\documentclass[fleqn,10pt]{wlscirep}
\begin{document}
\title{Photon statistics on the extreme entanglement}
 \author{Yang Zhang}
 \author{Jun Zhang}
 \author[1,*]{Chang-shui Yu}
\affil{School of Physics and Optoelectronic Technology, Dalian University of
Technology, Dalian 116024, China }
\affil[*]{quaninformation@sina.com;ycs@dlut.edu.cn}
\begin{abstract}
The effects of photon bunching and antibunching correspond to the classical and quantum features of the electromagnetic field, respectively. No direct evidence suggests whether these effects can be potentially related to quantum entanglement. Here we design a cavity quantum electrodynamics model with two atoms trapped in  to demonstrate the connections between the  steady-state photon statistics and the two-atom entanglement . It is found that within the weak dissipations and  to some good approximation, the
local maximal two-atom entanglements  perfectly correspond to not only the quantum feature of the electromagnetic field---the optimal
photon antibunching, but also  the classical feature---the optimal photon bunching. We also analyze the influence of strong dissipations and pure dephasing.  An intuitive physical understanding is also given finally.
\end{abstract}
\flushbottom
\maketitle
\thispagestyle{empty}
\section*{Introduction}

Nonlinear light-matter interaction is a long sought for quantum information
science \cite{11,22}, as well as a fascinating concept in terms of
fundamental physics. The strong interactions between individual photons is a
standing goal of both fundamental and technological significance \cite{lun}
. Photon blockade, as a typical nonlinear quantum optical effect, which
indicates the ability to control the nonlinear response of a system by the
injection of single photons \cite{Gerace,Dario,blocka}, shows that the system `blocks'
the absorption of a second photon with the same energy. The typical feature
is the photon antibunching which is signaled by a rise of $g^{(2)}(\tau )$
with $\tau $ increasing from $0$ to larger values while $g^{(2)}(0)<g^{(2)}(%
\tau )$ as discussed in detail in Ref \cite{Mandel,optimal, Carusotto}.  The converse
situation, $g^{(2)}(0)>g^{(2)}(\tau )$ is called photon bunching which  indicates large probability of more than one photon to arrive simultaneously to the detector. It is usually considered as a
purely classical behavior. As the peculiar feature of the quantum mechanical
nature, photon antibunching provides a way to controlling the single photon
via optical devices such as quantum optomechanical setups \cite%
{gong,Rabl,Nunnenkamp,87,liao}, feed back control system \cite{yulong},
superconducting circuit \cite{Hoffman,Liu}, quantum dots \cite{Winger,tang},
 Kerr-type nanostructured materials \cite{Dario}, confined cavity polaritons\cite{Verger}, cavity quantum electrodynamics (CQED) systems and so on  \cite{jf1,jf2,cio,wen,Liu,coherent,dd,Kimble,77,L,V,E,P,ss,smo}.

Recently,  the relation between photon statistics and other quantum effects have attracted increasing interests.  For example,  Ref. \cite{yulong,Liu} address the relation between photon blockade and optical bistability and Ref. \cite{Liu} also investigates the relation between photon blockade and  electromagnetically induced transparency.  In Ref.  \cite{wu} ,
 it is found how the photon blockade is affected by the parity-time symmetry. In addition,  the authors in Ref. \cite{oe}  find the connection between the first order correlation function and the violation of Bell inequalities.
As we know,  quantum entanglement is  not only an intriguing quantum feature but also
the important physical resource in quantum information processing\cite{li,yong,zhang}. Do there exist some relation between photon statistics and quantum entanglement? Or a weak question is whether one can design some particular quantum systems to create a potential relation.

 In this paper, we design a particular CQED model to reveal the relations between the photon statistics and atomic entanglement.
 Our model includes one
cavity weakly driven by a monochromatic laser field and two two-level atoms
trapped in the cavity. As mentioned above, photon statistics have been widely studied in CQED systems. Even though the mechanism of photon statistics is clear, intuitively, there is no proof that photon antibunching and bunching have any direct relation with entanglement.  So \textit{our interest is mainly to find the relation between
the photon statistics and the entanglement of the two atoms in a particular case instead of only
illustrating the photon statistics or atomic entanglement}. Firstly, we restrict our results in the weak dissipation regime and present our main result.  we  find that the maximal steady-state atomic entanglements as the quantum
feature just correspond to the quantum feature of the cavity field, that is,
the local optimal photon antibunching. It is surprising that the local
maximal steady-state atomic entanglements also perfectly correspond to the classical
feature of the field, that is, the photon bunching. However, the maximal
bunching point subject to a dark-state process corresponds to vanishing
steady-state entanglement. Secondly, we analyze the effects of strong dissipation as well as pure dephasing  on the correspondence relations. It is shown that entanglement is reduced faster than the second-order correlation function and the correspondences become worse and vanishing until the entanglement dies when the dissipations of the system are increased or the dephasing is considered.  We also discuss the experimental realization of our  proposal.  Finally, the intuitive physical analysis and some further discussions are provided.

\section*{Results}

\subsection*{ The physical model}

\begin{figure}[tbp]
\centering
\includegraphics[width=0.5\columnwidth,height=3in]{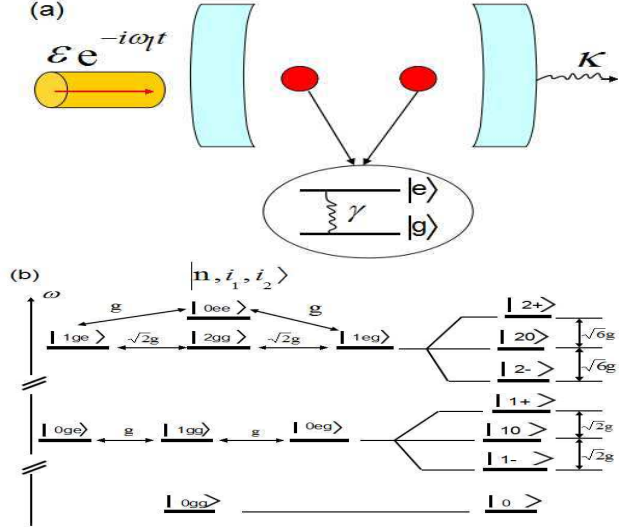} %
\caption{(color online). (a) A cavity coupled with two two-level atoms. The
driving field is weakly coupled to cavity mode with Rabi frequency $\protect%
\varepsilon $. $\protect\gamma $ and $\protect\kappa $ are the spontaneous
emission rate of the atoms and decay rate of the cavity, respectively. (b)
Energy levels corresponding to system's state up to $n_{ph}=2$. It indicates
the relevant transition processes between states and the possible excitation
pathways to state $\left\vert 2,g_{1},g_{2}\right\rangle $. The states are
labeled by $\left\vert n,i1,i2\right\rangle $ with $n$ denoting the photon
numbers of cavity mode, and $i1$ and $i2$ representing the levels of two
atoms, respectively. }
\end{figure}
As sketched in the Fig. 1(a), we study two two-level atoms coupled to the
cavity with frequency $\omega _{a}$ which is weakly driven by an external
optical field. The two-level atoms can be, in principle, replaced by any
two-level systems such as ions, quantum dots, superconductive qubits and so
on. The frequency of atomic transition from ground state $\left\vert
g\right\rangle $ to excited state $\left\vert e\right\rangle $ with
linewidth $\gamma $ is denoted by $\omega _{e}$. In this configuration \cite%
{Tan,nico} (we set $\hbar =1$ hereafter), the Hamiltonian can be given by
\begin{eqnarray}
H &=&\omega _{a}a^{\dag }a+\sum_{i=1}^{2}\omega _{e}\sigma _{i}^{+}\sigma
_{i}^{-}+\sum_{i=1}^{2}g_{i}(\sigma _{i}^{+}a+a^{\dag }\sigma _{i}^{-})
+\varepsilon \left( a^{\dag }e^{-i\omega _{L}t}+ae^{i\omega _{L}t}\right) ,
\label{(aa)}
\end{eqnarray}%
where $\sigma _{1,2}^{-}=\left\vert g_{1,2}\right\rangle \left\langle
e_{1,2}\right\vert $, $a(a^{\dag })$ are the annihilation (creation)
operators of the cavity mode and $g_{i}$ is the coupling coefficient between
the \textit{i}th atom and the cavity mode. The driving frequency is denoted
by $\omega _{L}$, and the driving strength by $\varepsilon $, respectively.
In the frame rotated at the laser frequency $\omega _{L}$, the Hamiltonian (%
\ref{(aa)}) becomes
\begin{eqnarray}
H &=&\Delta a^{\dag }a+\sum_{i=1}^{2}\delta \sigma _{i}^{+}\sigma
_{i}^{-}+\sum_{i=1}^{2}g_{i}(\sigma _{i}^{+}a+a^{\dag }\sigma _{i}^{-})
+\varepsilon \left( a^{\dag }+a\right) ,  \label{2}
\end{eqnarray}%
where $\Delta =\omega _{a}-\omega _{L}$ is the laser detuning from the
cavity mode and $\allowbreak \delta =\omega _{e}-\omega _{L}$ is the laser
detuning from the atoms.

For simplicity, here we assume that $g_{1}=g_{2}=g$ and we only consider
that the cavity is resonant with the atoms, i.e., $\omega_a=\omega_c$ and $%
\Delta=\delta$. Since the system is driven weakly, only few photons can be
excited. We can only focus on the few-photon subspace. Thus the Hamiltonian $%
H$ without driving can be easily diagonalized (we cut off the photons into
the two-photons subspace). The eigenvalues are given in the Methods and the
eigenstates in the current case, distinguished by different numbers of
photons, can be given as follows.
\begin{equation}
\left\vert 1_{0}\right\rangle =\frac{1}{\sqrt{2}}\left\vert
0,g,e\right\rangle -\frac{1}{\sqrt{2}}\left\vert 0,e,g\right\rangle ,
\end{equation}%
\begin{equation}
\left\vert 1_{+}\right\rangle =\frac{1}{\sqrt{2}}\left\vert
1,g,g\right\rangle +\frac{1}{2}\left\vert 0,g,e\right\rangle +\frac{1}{2}%
\left\vert 0,e,g\right\rangle ,
\end{equation}%
\begin{equation}
\left\vert 1_{-}\right\rangle =\frac{1}{\sqrt{2}}\left\vert
1,g,g\right\rangle -\frac{1}{2}\left\vert 0,g,e\right\rangle -\frac{1}{2}%
\left\vert 0,e,g\right\rangle ,
\end{equation}%
\begin{equation}
\left\vert 2_{01}\right\rangle =\frac{1}{\sqrt{3}}\left\vert
2,g,g\right\rangle -\frac{\sqrt{6}}{3}\left\vert 0,e,e\right\rangle ,
\end{equation}%
\begin{equation}
\left\vert 2_{02}\right\rangle =\frac{1}{\sqrt{2}}\left\vert
1,g,e\right\rangle -\frac{1}{\sqrt{2}}\left\vert 1,e,g\right\rangle ,
\end{equation}%
\begin{equation}
\left\vert 2_{+}\right\rangle =\frac{\sqrt{3}}{3}\left\vert
2,g,g\right\rangle +\frac{1}{2}\left\vert 1,g,e\right\rangle +\frac{1}{2}%
\left\vert 1,e,g\right\rangle +\frac{1}{\sqrt{6}}\left\vert
0,e,e\right\rangle ,
\end{equation}%
\begin{equation}
\left\vert 2_{-}\right\rangle =\frac{\sqrt{3}}{3}\left\vert
2,g,g\right\rangle -\frac{1}{2}\left\vert 1,g,e\right\rangle -\frac{1}{2}%
\left\vert 1,e,g\right\rangle +\frac{1}{\sqrt{6}}\left\vert
0,e,e\right\rangle .
\end{equation}
The energy levels are shown in the Fig. 1(b). The nonlinearity in the
coupling between the atoms and cavity gives rise to energy level structure
which can exhibit bunching and antibunching behaviors due to the splitting
of the eigen-energy \cite{jc}. With this structure, we can get an intuitive
picture for the different photon statistics.

\subsection*{ The photon statistics}
\begin{figure}[tbp]
\centering
 \includegraphics[width=0.8\columnwidth,height=4in]{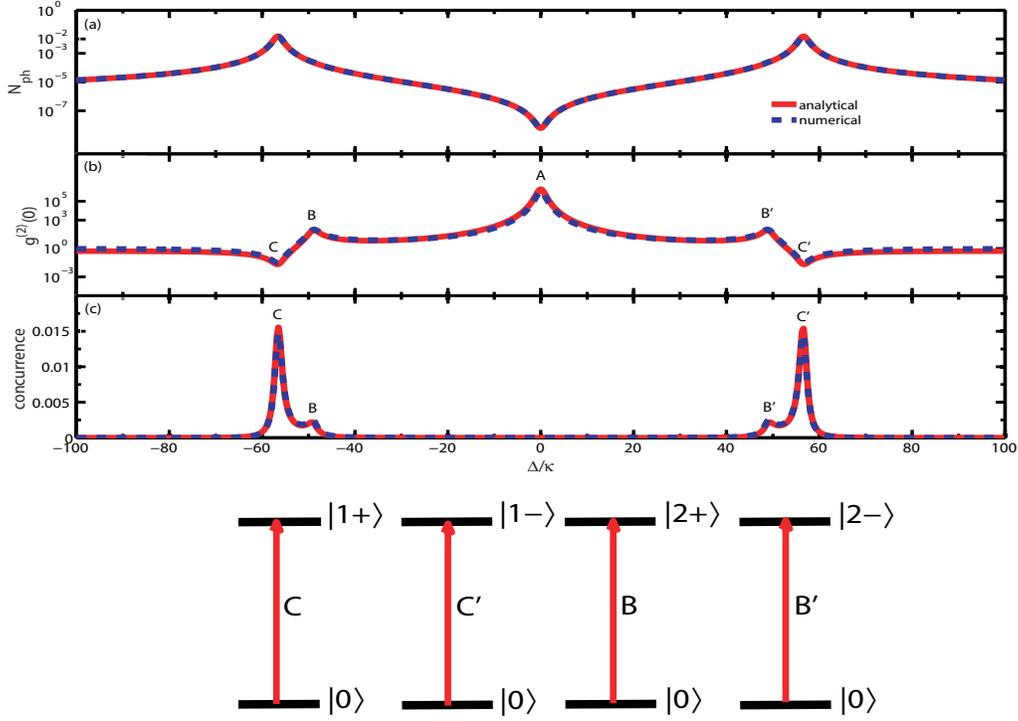}
\caption{(color online). The mean photon numbers of the cavity, the
equal-time second-order function $g^{(2)}(0)$ and the entanglement of two
atoms vs the detuning $\Delta $, respectively. The red curves are
approximate and analytical solution of Eq. (\ref{8eq}), Eq. (\ref{aa}) and Eq. (\ref{bb}). The
blue curves are numerical results of the quantum master equation Eq. (\ref{master}) . We
take $\protect\gamma /\protect\kappa =1,$ $g/\protect\kappa =40$, $\protect%
\varepsilon /\protect\kappa =0.125$.  At the bottom, we also show the transition between the different eigenstates corresponding to the different points.}
\end{figure}
As mentioned at the beginning, the different photon statistics are signaled
by the equal-time (namely zero-time-delay) second-order photon-photon
correlation function \cite{scully} which reads
\begin{equation}
g^{(2)}(0)=\frac{\left\langle a^{\dagger }a^{\dagger }aa\right\rangle }{%
\langle a^{\dagger }a\rangle ^{2}}=\frac{\sum_{n}n(n-1)p_{n}}{%
(\sum_{n}np_{n})^{2}},  \label{nn}
\end{equation}%
where $n=\langle a^{\dagger }a\rangle $ is the intra-cavity photon number of
the cavity mode, $p_{n}$ represents the probability with $n$ photons. In Eq.
(\ref{nn}) the operator is evaluated at the same time. When the second-order
correlation function satisfies the inequality $g^{(2)}(0)\leq 1$, there
occurs the photon antibunching, i.e., the photon blockade which means the
system 'blocks' the absorption of a second photon with the same energy with
large probability. The limit $g^{(2)}(0)\rightarrow 0$ means the perfect
photon blockade in which two photons never occupy the cavity at the same
time. On the contrary, when $g^{(2)}(0)>1$, it means that photons inside the
cavity enhance the resonantly entering probability of subsequent photons
\cite{xuxunwei,li1,li2,lang} .

To give an intuitive picture and gain more insight into the physics, we
first take an analytic (but approximate) method to calculate the
second-order correlation function by employing the wave function amplitude
approach. Considering the effects of the leakage of the cavity $\kappa $,
the spontaneous emission $\gamma $ of the atoms, we phenomenologically add
the relevant damping contributions to Eq. (\ref{2}). Thus the Hamiltonian can
be rewritten as $H-\frac{i}{2}(\kappa a^{\dagger }a+\gamma
\sum_{i=1}^{2}\sigma _{i}^{+}\sigma _{i}^{-})$. Analogous to the above
statements, the photon number is up to 2. So one can assume that the state
of the composite system is given by \cite{fliuds,steady1}
\begin{eqnarray}
\left\vert \Psi \right\rangle &=&A_{0gg}\left\vert 0,g,g\right\rangle
+A_{0ge}\left\vert 0,g,e\right\rangle +A_{0eg}\left\vert 0,e,g\right\rangle
+A_{1gg}\left\vert 1,g,g\right\rangle +A_{1ge}\left\vert 1,g,e\right\rangle
+A_{1eg}\left\vert 1,e,g\right\rangle  \notag \\
&&+A_{0ee}\left\vert 0,e,e\right\rangle +A_{2gg}\left\vert
2,g,g\right\rangle .  \label{phi}
\end{eqnarray}%

So the dynamical evolution of the state Eq. (\ref{phi}) subject to the
damping Hamiltonian is given by
\begin{equation}
i\dot{A}_{0eg}=(\Delta -i\gamma /2)A_{0eg}+gA_{1gg}+\varepsilon A_{1eg},
\end{equation}\label{hh1}
\begin{equation}
i\dot{A}_{0ge}=(\Delta -i\gamma /2)A_{0ge}+gA_{1gg}+\varepsilon A_{1ge},
\end{equation}%
\begin{equation}
i\dot{A}_{1gg}=(\Delta -i\kappa /2)A_{1gg}+g(A_{0eg}+A_{0ge})+\sqrt{2}%
\varepsilon A_{2gg}+\varepsilon A_{0gg},
\end{equation}%
\begin{equation}
i\dot{A}_{1eg}=(2\Delta -i\gamma /2-i\kappa )A_{1eg}+gA_{0ee}+\sqrt{2}%
gA_{2gg}+\varepsilon A_{0eg},
\end{equation}%
\begin{equation}
i\dot{A}_{0ee}=2(\Delta -i\gamma /2)A_{0ee}+g(A_{1eg}+A_{1eg}),
\end{equation}%
\begin{equation}
i\dot{A}_{1ge}=(2\Delta -i\gamma /2-i\kappa /2)A_{1ge}+gA_{0ee}+\sqrt{2}%
gA_{2gg}+\varepsilon A_{0ge},
\end{equation}%
\begin{equation}
i\dot{A}_{2gg}=2(\Delta -i\kappa /2)A_{2gg}+\sqrt{2}g(A_{1eg}+A_{1ge})+%
\varepsilon A_{1gg}.
\end{equation}\label{hh2}
From Eq. (\ref{phi}), one can easily write $g^{(2)}(0)=\frac{2p_{2}}{%
(p_{1}+2p_{2})^{2}}$ with $p_{1}=\left\vert \bar{A}_{1gg}\right\vert
^{2},p_{2}=\left\vert \bar{A}_{2gg}\right\vert ^{2}$. Since the weakly
driving is considered, one can easily get $p_{1}\gg $ $p_{2}$, which means
that $g^{(2)}(0)$ can be simplified as $g^{(2)}(0)=\frac{2p_{2}}{p_{1}^{2}}$%
. Under the stability conditions, we can easily obtain the steady-state
solution of Eqs. (12-18) by letting the derivatives on the
left-hand-side vanish. The concrete expressions of the steady solutions are
given in the Methods. Substitute the steady-state solution into $g^{2}(0)$,
one can immediately arrive at
\begin{equation}
g^{(2)}(0)=\frac{2\left\vert \bar{A}_{2gg}\right\vert ^{2}}{\left\vert \bar{A%
}_{1gg}\right\vert ^{4}}=\left\vert \frac{(l+m)x}{2(y+z)(\Delta -i\gamma
)^{2}}\right\vert ^{2}.  \label{aa}
\end{equation}%
In addition, the mean photon number can also be given by
\begin{equation}
N_{ph}=\left\langle \Psi \right\vert a^{\dagger }a\left\vert \Psi
\right\rangle =\left\vert \frac{(\Delta -i\gamma )\varepsilon }{x}%
\right\vert ^{2}.  \label{8eq}
\end{equation}%
Note that here
\begin{eqnarray}
x &=&2g^{2}-(\Delta -i\gamma /2)(\Delta -i\kappa /2), \\
y &=&(\gamma /2+i\Delta )(\Delta -i\kappa /2)(\gamma /2+2i\Delta )+\kappa
/2),  \notag \\
z &=&g^{2}(3\Delta -i\gamma -i\kappa /2),  \notag \\
l &=&-ig^{2}(1-2\sqrt{2})(\gamma /2+i\Delta ))),  \notag \\
m &=&(\gamma /2+i\Delta )^{2}(\gamma /2+2i\Delta +\kappa /2).  \notag
\end{eqnarray}
In order to show the validity of the above analytic treatment, we also
employ the quantum master equation to numerically study the above results.
Considering the above quantum system, the Markovian quantum master equation
reads
\begin{eqnarray}
\dot{\rho}=-i[H,\rho ]+\frac{\kappa }{2}(2a\rho a^{\dagger }-a^{\dagger
}a\rho -\rho a^{\dagger }a)+\sum_{i=1}^{2}\frac{\gamma }{2}(2\sigma
_{i}^{-}\rho \sigma _{i}^{+}-\sigma _{i}^{+}\sigma _{i}^{-}\rho -\rho \sigma
_{i}^{+}\sigma _{i}^{-}),  \label{master}
\end{eqnarray}
where $H$ is the Hamiltonian given by Eq. (\ref{2}), $\rho $ is the density
operator of the whole composite system, and $L[\hat{d}]\rho =2\hat{d}\rho
\hat{d}^{\dagger }-\hat{d}^{\dagger }\hat{d}\rho -\rho \hat{d}^{\dagger }%
\hat{d},(\hat{d}=\hat{a},\sigma _{i}^{-},i=1,2)$ is the dissipator. In
addition, we don't consider the thermal photons for simplicity \cite%
{xuxunwei} . Since the steady-state solution is needed for our purpose, we
will directly employ a numerical way to solving Eq. (\ref{master}) for the
steady state $\rho _{s}$ \cite{tan1} . So the second-order correlation
function can be directly obtained by $g^{(2)}(0)=\frac{Tr[\rho
_{s}a^{\dagger 2}a^{2}]}{[Tr(\rho _{s}a^{\dagger }a)]^{2}}$ and the mean
photon number is obtained by $N_{ph}=$ $Tr(\rho _{s}a^{\dagger }a)$.
\begin{figure}[tbp]
\centering
\hspace*{0cm}\includegraphics[width=0.5\columnwidth,height=2.5in]{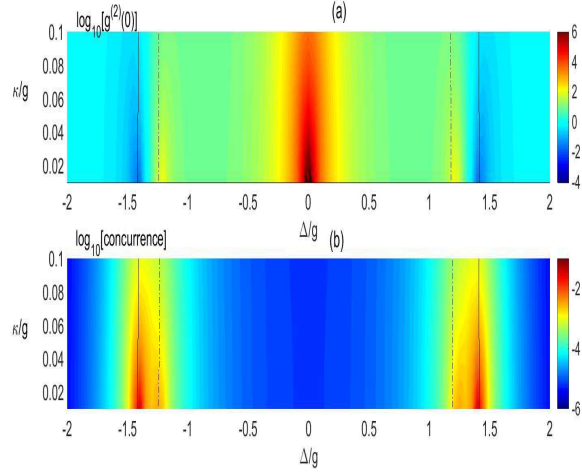}
\caption{(color online). We plot logarithm of the equal-time second-order
function $\log _{10}\left[ g^{(2)}(0)\right] $ and the logarithm of
concurrence $\log _{10}\left[ concurrence\right] $ as a function of the
detuning $\Delta $ and cavity decay rate $\protect\kappa $. (a) shows the
second-order correlation function and (b) corresponds to the entanglement of
atoms. The locally optimal photon antibunching and bunching and the locally
maximal entanglement are also illustrated by the black-solid line and
black-dashed line respectively in (a) and (b) which corresponds to $\Delta
^{2}=2g^{2}$ and $\Delta ^{2}=3/2g^{2}$, respectively. Here, we set $\protect%
\gamma =\protect\kappa $, $\protect\varepsilon /g=0.0065$.}
\end{figure}

In Fig. 2 (a) and Fig. 2 (b), we plot the mean cavity numbers $N_{ph}$ and
second-order correlation function $g^{(2)}(0)$ changing with the detuning $%
\Delta$ in the case of weak dissipations. One can find that the numerical
results given by Eq. (\ref{master}) and the analytic and approximate results
given by Eqs. (\ref{aa}) and (\ref{8eq}) show the perfect agreement. This
guarantees that all the following conclusions drawn from our analytical way
is valid. Let's focus on Fig. 2 (a) and (b). It is shown that the points $C$
and $C^\prime$ ($g^{(2)}(0)\ll 1$) where photon statistics satisfy the
sub-Poissonian distribution, correspond to the photon blockades which are
the local optimal antibunching points in this system. At these two points, $%
\Delta =\pm \sqrt{2}g$ which means that the driving field is just resonant
with the transition between the single-photon polariton states $\left\vert
1,\pm \right\rangle $ and the ground state $\left\vert 0\right\rangle$. In
this case, once the first photon excited the transition from $\left\vert
0\right\rangle$ to $\left\vert 1,\pm \right\rangle $ by the coherent
driving, the photon with the same energy is not resonant with any other
transition (the energy does not match between any other two levels). So it
seems that the first photon 'blocks' the absorption of a second photon. At
points $B$ and $B^\prime$, one can find $g^{(2)}(0)> 1$ which corresponds to
the photon bunching. At these two points, $\Delta =\pm \frac{\sqrt{6}}{2}g$
which correspond to the resonance between the driving field and the
transition from the ground state $\left\vert 0\right\rangle$ to the excited
states $\left\vert 2,\pm \right\rangle $. It indicates a resonance process
of double photons. At point $A$, it shows a strong photon bunching effect
with $g^{(2)}(0)\gg 1$ . At this point, $\Delta =0$ and the photon
statistics satisfy the super-Poissonian distribution. This does not
correspond to a resonance process. When $\Delta=0$, the system is coherently
driven into a dark state $\left\vert dark\right\rangle =g\left\vert
0,g,g\right\rangle -\frac{\varepsilon }{2}(\left\vert 0,g,e\right\rangle
+\left\vert 0,e,g\right\rangle )$. The state $\frac{1}{\sqrt{2}}(\left\vert
0,g,e\right\rangle -\left\vert 0,e,g\right\rangle )$ is allowed to transit
to $\frac{1 }{\sqrt{2}}(\left\vert 1,g,e\right\rangle -\left\vert
1,e,g\right\rangle )$ which is strongly coupled to the state $\left\vert
2gg\right\rangle$. This is similar to the electromagnetically induced
transparency \cite{E1,E2} .

\subsection*{Atomic entanglement and photon statistics}

Since we have calculated the state $\left\vert\Psi\right\rangle$ given in
Eq. (\ref{phi}), one can easily obtain the reduced density matrix $\rho_{AB}$
for the two atoms. Thus one can also easily calculate the corresponding
entanglement. Here in order to show the two-atom entanglement, we would like
to employ Wootters' concurrence as the entanglement measure \cite{wooter}
which, for the bipartite density matrix of qubits, is defined by
\begin{equation}
C(\rho _{AB})=\max \{0,\sqrt{\lambda _{1}}-\sqrt{\lambda _{2}}-\sqrt{\lambda
_{3}}-\sqrt{\lambda _{4}}\},  \label{cc}
\end{equation}%
where $\lambda _{i}$ is the square root of the $i$th eigenvalue of the $\rho
_{AB}\tilde{\rho}_{AB}$ in decreasing order with $\tilde{\rho}_{AB}=(\sigma
_{y}\otimes \sigma _{y})\rho _{AB}^{\ast }(\sigma _{y}\otimes \sigma _{y})$.
Substituting $\rho _{AB}$ in to Eq. (\ref{cc}) one can easily obtain (see
Methods)
\begin{equation}
C(\rho _{AB})=2\left\vert A_{0ee}-A_{0ge}A_{0eg}\right\vert .  \label{con}
\end{equation}%
Note that in Eq. (\ref{con}), we have neglected the terms with the power of $%
\varepsilon$ more than 2. Based on the steady amplitudes derived from Eq.
(6), the concurrence is analytically given by
\begin{equation}
C(\rho _{AB})=\left\vert \frac{g^{2}\varepsilon ^{2}(Gx-2(y+z))}{2x^{2}(y+z)}%
\right\vert ,  \label{bb}
\end{equation}%
with $G=\sqrt{2}(\gamma +i\Delta +\sqrt{2}i\Delta +\sqrt{2}\kappa )$.

We have plotted Eq. (\ref{bb}) in Fig. 2 (c). The concurrence via the
numerical way (by solving Eq. (\ref{master})) is also plotted in this
figure. One can find that the analytic concurrence matches the numerical
results very well, which guarantees the validity of our approximate and analytic
results. Although the steady-state concurrence is not so large in contrast to Ref. \cite{steadye} which is essentially within a different mechanism (two coupled and driven cavities and  relatively large driving-dissipation ratio $\sim10^3$), it does not affect our purpose of this paper. From Fig. 2 (c), it is obvious that the concurrence has two pairs
of local maximal values. Compared with Fig. 2 (b), one can easily find that
these two local maximal entanglement perfectly correspond to the local
optimal photon antibunching and bunching. Such a correspondence can also be
supported by the analytic expression given in Eq. (\ref{bb}), from which one
can see that the extrema occur at $\Delta ^{2}=2g^{2}$ and $\Delta ^{2}=%
\frac{3}{2}g^{2} $ for small $\{\kappa ,\gamma \}$. This is consistent with
the above analysis on the photon statistics. Next we will provide give a relatively intuitive
understanding of this correspondence. One should first note from Eq. (\ref%
{con}) that only the three parameters $A_{0ee}$, $A_{0eg}$ and $A_{0ge}$
play the dominant role in entanglement. So at $\Delta ^{2}=2g^{2}$, the
driving field is tuned resonantly with the transition between $\left\vert
0gg\right\rangle $ and $\left\vert 1,\pm \right\rangle $ which leads to the
optimal photon blockade. In addition, the strength of such a resonant
interaction is proportional to the first order of the driving field $%
\varepsilon $. So $\left\vert 1,\pm \right\rangle $ gets a relatively large
proportion in the total state $\left\vert \Psi \right\rangle $. It is
obvious from Eq. (\ref{con}) or Eq. (\ref{bb}) that $\left\vert \Psi
\right\rangle $ owns the relatively large amount of entanglement ($C$ and $%
C^{\prime }$ in Fig. 2 (c)). If $\Delta ^{2}=\frac{3}{2}g^{2}$, the driving
field is resonant with the transition between $\left\vert 0gg\right\rangle $
and $\left\vert 2,\pm \right\rangle $. Thus $\left\vert 2,\pm \right\rangle $
occupies the relatively dominant proportion in $\left\vert \Psi
\right\rangle $. However, the interaction strength is proportional to the
second order of $\varepsilon ^{2}$. So the entanglements at these points get
the extremum ($B$ and $B^{\prime }$ ), but they are still much less than the
entanglement at $C$ and $C^{\prime }$. We would like to point that the consistency between photon statistics and entanglement is attribute to that they  can be understood in a unified and intuitive way.  The resonant transitions of both single-photon process and
double-photon process, as the essential physics of photon antibunching and bunching, correspond to the superposition of the ground state
and an entangled  state.  So the maximal atomic entanglement is well
consistent with photon statistics. Given the (weak) driving strength, the
single-photon process happens with a much larger probability than of
double-photon process, so the entanglement subject to double-photon process
is small. However, it is interesting that the
maximal photon bunching point at $\Delta =0$ does not correspond to an
extremum of entanglement. The reason is attributed to the dark-state process
which provides a channel (as mentioned in the part of photon statistics) to
be converted to the state $\left\vert 2gg\right\rangle $ as well as $%
\left\vert 0ee\right\rangle $. Their proportions in $\left\vert \Psi
\right\rangle $ get relatively larger. The net effect on entanglement is
that $\left\vert 0ee\right\rangle $ and $\left\vert 0ge\right\rangle $, $%
\left\vert 0eg\right\rangle $ reach a balance subject to Eq. (\ref{con}), so
the entanglement is negligibly small.
\begin{figure}[tbp]
\centering
\hspace*{0cm}
\includegraphics[width=1\columnwidth,height=3in]{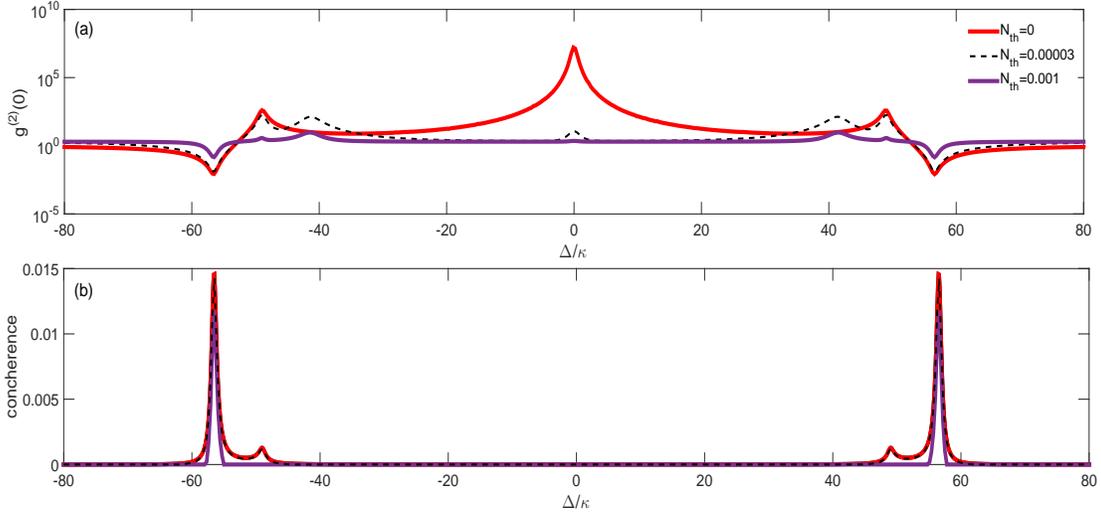}
\caption{(color online).  $g^{(2)}(0) $ and the concurrence versus the
detuning with different $\bar{N}_{th}$. All the parameters are the same as Fig. 2. The figure shows how the correspondence is gradually destroyed by thermal environments. }
\end{figure}

We would like to emphasize that all our presented correspondence relations hold within the weak dissipation regime. Once this condition is not satisfied, these relations will be reduced or destroyed. In order to show the influences, we first plot the concurrence and $g^{2}(0)$ via $\gamma
=\kappa $ and $\Delta $ in Fig. 3 in the vacuum environments. One can find that both the local optimal
photon statistics and the concurrence extrema are reduced with $\kappa $
increasing. Meanwhile, the correspondence relation between concurrence and $%
g^{2}(0)$ gets a little bit worse. This can be well understood from Eq. (\ref%
{aa}) and Eq. (\ref{bb}) from which one can see that all the relevant
analysis are satisfied within the error region to the same order as $\kappa
^{2}$ (we assume $\gamma =\kappa $ for simplicity). So we always
limit our study in the region with small enough dissipations. Physically,
the large deviation of the correspondence is directly attributed to the
large line width of the level induced by the dissipations.

 Next, we will consider how thermal photons and the atomic dephasing influence our results. In fact, it can be easily predicted that our results will be destroyed greatly since quantum feature (especially the entanglement) is generally quite fragile to these environments. We consider the thermal environments by solving the following master equation $\dot{\rho}=-i[H,\rho ]+\frac{\kappa }{2}(\bar{N}%
_{th}+1)L[a]+\frac{\kappa }{2}\bar{N}_{th}L[a^{\dagger }]+\sum_{i=1}^{2}%
\frac{\gamma }{2}(\bar{N}_{th}+1)L[\sigma _{i}^{-}]+\sum_{i=1}^{2}\frac{%
\gamma }{2}\bar{N}_{th}L[\sigma _{i}^{+}]$ where $\bar{N}_{th}=[\exp (\hbar \nu /k_{B}T)-1]^{-1}$ is the average photon number with  $k_{B}$ denoting the
Boltzmann constant and $T$ standing for the reservoir temperature. In addition, all the other parameters are defined the same as Eq. (\ref{master}). The numerical results are shown in Fig. 4 where we  can observe that with the increasing of $\bar{N}_{th}$, the entanglement, photon antibunching and photon bunching are all reduced at the correspondence points, but the entanglement decays very fast and even dies with large  $\bar{N}_{th}$. But the correspondence relations can  be kept with less thermal photons until the entanglement vanishes. In addition,  photon bunching can be enhanced at other places, which just shows the participation of the thermal photons. We also consider the effect of atomic dephasing procedure\cite{dephasing}, which is done by adding a Lindblad term $\gamma_{d}L[\sigma _{i}^{+}\sigma
_{i}^{-}](i=1,2)$ in Eq. (\ref{master}).  Numerical procedure shows the completely similar results as the thermal environment. So the figures are omitted here.
\section*{Experimental realization }

Up to now, based on the schematic setup in our paper,  we have theoretically studied the  relation between photon blockade and the atomic entanglement and presented the physics behind this scheme. In the following, we will give a brief 
analysis on whether the conditions that we require are achievable in practical experiments.   Based on the  previous sections, we should note that  the possibility to realize the proposal mainly depends on the strong coupling rate ( $g/\kappa\sim 40$ and $\gamma/\kappa\sim 1$ for our numerical simulation). Thus, we will extensively focus on the parameters  $g$, $\kappa$ and $\gamma$. As mentioned above, our physical model is not restricted in the real atomic systems. Let's consider the quantum device circuit QED system (circuit QED-consisting of microwave resonators and superconducting qubit) \cite{Fink,Alexandre} or  quantum dot coupled with the photonic crystal cavity \cite{ande}. In circuit QED system, the strong coupling can be realized and the long coherence time of a superconducting qubit embedded in a high-quality on-chip microwave cavity \cite{Fink}. The cavity-qubit coupling strengths can be realized experimentally from $2\protect\pi\times5.8 $ MHz to $g_{max}=2\protect\pi\times 210$ MHz and the relaxation time of the qubit can reach $7.3\mu s$ \cite{Schuster,Wallraf} which corresponds to the decay rate $\gamma\sim{2\pi}\times 0.02$ MHz. The qubit transition frequencies can be chosen anywhere from about 5 GHz to 15 GHz \cite{Alexandre} and can be tuned by applying a magnetic flux through the qubit loop. The cavity decay rate $\kappa$ can be as low as $2\protect\pi\times 5$ KHz due to the high value of the quality factor Q with resonator frequency to be  between $5$ GHz and $10$ GHz \cite{Alexandre,Peter}.  So the ratio used in our simulations $g/\protect\kappa =40$ and $\gamma\sim\kappa$ are reasonable and easily achieved.  In addition, the system can be cooled  to temperatures below 20 mK \cite{Fink,science} (15 mK in \cite{Bishop}) in a dilution refrigerator. Correspondingly, the number of thermal photon $\bar{N}_{th}$ subject to the transition frequency $2\pi\times 6.5$ GHz for the qubit is  less than $\bar{N}^\prime_{th}=1.66\times 10^{-7}$ (for 20 mK). It can even be adjusted to  $\bar{N}^\star_{th}=1.23\times 10^{-14}$ (for 15 mK and transition frequency $2\pi\times 10$ GHz ). From Fig. 4, one can find that the entanglement is hardly affected by $\bar{N}_{th}=0.00003$, even though it is usually fragile for noise. It can be reasonably predicted that if $\bar{N}_{th}\rightarrow\bar{N}^\prime$, even $\bar{N}^\star_{th}$, our correspondence relation will be perfectly observed in experiment. The dephasing of the qubit in one realization of this system has also been measured in Ref \cite{Wallraf11,Schreier}. It shows that the pure dephasing time $T_{d} $  can reach as long as $5.5 \mu s$ \cite{Schreier} which translates to $\gamma_{d}=2\protect\pi\times0.03 $ MHz $\approx1.5\times 10^{-4} g_{max}$. We can loosely choose $\kappa$ and $\gamma$ such that $g/40\sim\kappa\sim \gamma\gg\gamma_{d}$ is achieved,  so the effect of dephasing can be safely omitted here \cite{Ginossar}. Based on the above analysis, one can easily find that  all the conditions required for the demonstration of the correspondence relation are realizable within the current experimental technology.

\section*{Discussion}

To sum up, we have analyzed the physical mechanisms of photon statistics and
entanglement in detail. We find that the local maximal entanglement always
correspond to the local optimal photon bunching and antibunching points. In
other words, the local extremum of photon statistics subject to the
resonance processes are in good agreement with the local maximal
entanglement. However, the maximal photon bunching point corresponds to the
almost vanishing entanglement due to the dark-state process. One could think
that the correspondence between atomic entanglement and photon antibunching
could be easily understood since both of them are the quantum feature,
whereas it could be strange that the quantum feature (atomic entanglement)
corresponded to a classical effect (photon bunching). We also consider how the correspondence is affected by thermal noises and pure dephasing.

In addition, we would like to provide a qualitative physical interpretation again.  In CQED model, the photon antibunching  essentially corresponds to  resonant transition between the ground state and the single-excitation eigen-modes and bunching corresponds to the transition between the ground state and the two-excitation eigen-modes. Once such transitions happen, the trapped double atoms have $50\%$ probability to only absorb one photon to form a maximally-entangled-state component in the corresponding eigen-mode. This is the key matching mechanism. So photon statistics corresponding to such transition procedures are consistent to the extremum entanglement. But the double-excitation procedures happen with relatively little probability due to the weak driving, so the entanglement is much smaller. All the above analysis are obviously limited under the condition that the incoming photon (energy) can be well kept and no extra photons disturb this matching mechanism. This just means the weak dissipation. On the contrary, the strong decays ($\kappa$ and $\gamma$), the large thermal photon number as well as the dephasing reduce and even break the matching relation, so the correspondence gets worse. The dark-state process is another path which reaches the photon bunching around the mentioned matching mechanism, so there is no entanglement at this point. Therefore, we emphasize that the correspondence should be taken into account within weak dissipations. The proposal is within reach by current technologies, especially in the state-of-the-art circuit QED system.

Finally, we want to say that there are other relevant questions deserving us forthcoming efforts. For example, is there other mechanism leading to such a correspondence, or can we find other models with stronger correspondence? Can we effectively use this relation to control photon statistics by  entanglement, or on the contrary, to control entanglement by photon statistics?

\section*{Methods}

\textbf{{Eigenvalues of Hamiltonian.}-} The Hamiltonian without driving in
Eq. (\ref{2}) can be easily diagonalized in few-photon subspace. For
integrity, here we would like to provide the concrete expressions of the
eigenvalues of the Hamiltonian. Note that the driving frequency is retained
and we cut off the photon number up to 2.
\begin{equation}
\begin{aligned} E_{1_{a}}=\delta,\\ E_{1_{b}}=\Delta +\sqrt{2g},\\
E_{1_{c}}=\Delta -\sqrt{2g},\\ E_{2_{a}}=2\Delta,\\ E_{2_{b}}=2\Delta
+\sqrt{6g},\\ E_{2_{c}}=2\Delta -\sqrt{6g}.\\ \end{aligned}
\end{equation}
\textbf{{The steady-state solution of Eq.(12-18).}- }In order to
obtain the steady-state solution of Eq. (12)-Eq. (18), we set the time
derivatives to be zero and solve the equations within the weak driving
limit. We assume $\bar{A}_{0gg}\rightarrow1$, and drop the terms of the
power of $\varepsilon$ more than 2. The solutions are given as follows.
\begin{equation}
\bar{A}_{0eg}=-\frac{g\varepsilon }{2g^{2}+(\gamma +i\Delta )(\kappa
+i\Delta )},
\end{equation}
\begin{equation}
\bar{A}_{0ge}=-\frac{g\varepsilon }{2g^{2}+(\gamma +i\Delta )(\kappa
+i\Delta )},
\end{equation}
\begin{equation}
\bar{A}_{1gg}=-\frac{i\varepsilon (\gamma +i\Delta )}{2g^{2}+(\gamma
+i\Delta )(\kappa +i\Delta )},
\end{equation}%
\begin{equation}
\bar{A}_{1eg}=\frac{i g \varepsilon ^2 (\gamma +i \Delta ) \left(\sqrt{2}
\gamma +\left(2+\sqrt{2}\right) i \Delta +2 \kappa \right)}{2 \left(2
g^2+(\gamma +i \Delta ) (\kappa +i \Delta )\right) \left((\gamma +i \Delta )
(\kappa +i \Delta ) (\gamma +2 i \Delta +\kappa )+g^2 (2 \gamma +3 i \Delta
+\kappa )\right)},
\end{equation}
\begin{equation}
\bar{A}_{1ge}=\frac{i g \varepsilon ^2 (\gamma +i \Delta ) \left(\sqrt{2}
\gamma +\left(2+\sqrt{2}\right) i \Delta +2 \kappa \right)}{2 \left(2
g^2+(\gamma +i \Delta ) (\kappa +i \Delta )\right) \left((\gamma +i \Delta )
(\kappa +i \Delta ) (\gamma +2 i \Delta +\kappa )+g^2 (2 \gamma +3 i \Delta
+\kappa )\right)},
\end{equation}%
\begin{equation}
\bar{A}_{0ee}=\frac{g^2 \varepsilon ^2 \left(\sqrt{2} \gamma +\left(2+\sqrt{2%
}\right) i \Delta +2 \kappa \right)}{2 \left(2 g^2+(\gamma +i \Delta )
(\kappa +i \Delta )\right) \left((\gamma +i \Delta ) (\kappa +i \Delta )
(\gamma +2 i \Delta +\kappa )+g^2 (2 \gamma +3 i \Delta +\kappa )\right)},
\end{equation}
\begin{equation}
\bar{A}_{2gg}=-\frac{\varepsilon ^2 (\gamma +i \Delta ) \left(\left(1-2
\sqrt{2}\right) g^2+(\gamma +i \Delta ) (\gamma +2 i \Delta +\kappa )\right)%
}{2 \left(2 g^2+(\gamma +i \Delta ) (\kappa +i \Delta )\right) \left((\gamma
+i \Delta ) (\kappa +i \Delta ) (\gamma +2 i \Delta +\kappa )+g^2 (2 \gamma
+3 i \Delta +\kappa )\right)}.
\end{equation}
\textbf{Concurrence of the two atoms.-} Here we give a detailed
derivation of the concurrence given in Eq. (\ref{con}). Since we have
obtained the steady-state solution of Eq. (12)-Eq. (18), one can find that
the photon number and the excitation number in the subscript of $A_{ijk}$
signal the power of $\varepsilon$ in $A_{ijk}$. From the state $%
\left\vert\Psi\right\rangle$, one can find that the reduced density matrix of
the two atoms can be given by
\begin{equation}
\rho_{AB}=Tr_{C}\left\vert\Psi\right\rangle\left\langle\Psi\right\vert=\left%
\vert\psi_0\right\rangle\left\langle\psi_0\right\vert+\left\vert\psi_1\right%
\rangle\left\langle\psi_1\right\vert+\left\vert\bar{A}_{2gg}\right\vert^2%
\left\vert gg\right\rangle\left\langle gg\right\vert,  \label{red}
\end{equation}
where
\begin{equation}
\left\vert\psi_0\right\rangle=\bar{A}_{0gg}\left\vert gg\right\rangle+\bar{A}%
_{0ge}\left\vert ge\right\rangle+\bar{A}_{0eg}\left\vert eg\right\rangle+%
\bar{A}_{0ee}\left\vert ee\right\rangle
\end{equation}
\begin{equation}
\left\vert\psi_1\right\rangle=\bar{A}_{1gg}\left\vert gg\right\rangle+\bar{A}%
_{1ge}\left\vert ge\right\rangle+\bar{A}_{1eg}\left\vert eg\right\rangle+%
\bar{A}_{1ee}\left\vert ee\right\rangle
\end{equation}
and the subscript $C$ means trace over cavity field. In order to calculate
the concurrence defined by Eq. (\ref{cc}), we need to calculate the matrix $%
\rho S\rho^*S$ with $S=\sigma_y\otimes\sigma_y$. Thus one can have
\begin{equation}
\rho S\rho^*S=\left\vert\psi_0\right\rangle\left\langle\psi_0\right\vert
S\left\vert\psi^*_0\right\rangle\left\langle\psi^*_0\right\vert S+M,
\end{equation}%
with
\begin{eqnarray}
M&=&\left\vert\psi_0\right\rangle\left\langle\psi_0\right\vert
S\left\vert\psi^*_1\right\rangle\left\langle\psi^*_1\right\vert S +\left\vert%
\bar{A}_{2gg}\right\vert^2\left\vert\psi_0\right\rangle\left\langle\psi_0%
\right\vert S\left\vert gg\right\rangle\left\langle gg\right\vert S
+\left\vert\psi_1\right\rangle\left\langle\psi_1\right\vert
S\left\vert\psi^*_0\right\rangle\left\langle\psi^*_0\right\vert
S+\left\vert\psi_1\right\rangle\left\langle\psi_1\right\vert
S\left\vert\psi^*_1\right\rangle\left\langle\psi^*_1\right\vert S  \notag \\
&&+\left\vert\bar{A}_{2gg}\right\vert^2\left\vert\psi_1\right\rangle\left%
\langle\psi_1\right\vert S\left\vert gg\right\rangle\left\langle
gg\right\vert S+\left\vert\bar{A}_{2gg}\right\vert^2\left\vert
gg\right\rangle\left\langle gg\right\vert S\left\vert
\psi^*_0\right\rangle\left\langle \psi^*_0\right\vert S +\left\vert\bar{A}%
_{2gg}\right\vert^2\left\vert gg\right\rangle\left\langle gg\right\vert
S\left\vert \psi^*_1\right\rangle\left\langle \psi^*_1\right\vert S.
\end{eqnarray}
Since $\bar{A}_{0gg}\rightarrow 1$, $\left\vert\psi_0\right\rangle\left%
\langle\psi_0\right\vert
S\left\vert\psi^*_0\right\rangle\left\langle\psi^*_0\right\vert
S\sim\varepsilon^0$ and $M\sim\varepsilon^2$. Thus $M$ can be regarded as
the perturbation. To proceed, we can find that the eigenvalue and the left
and right eigenvectors of the matrix $\left\vert\psi_0\right\rangle\left%
\langle\psi_0\right\vert
S\left\vert\psi^*_0\right\rangle\left\langle\psi^*_0\right\vert S$ are $%
C^2\left(\left\vert\psi_0\right\rangle\right)$, $\left\langle
\psi_0^*\right\vert S$ and $\left\vert\psi_0\right\rangle$, respectively. So
the 'first-order' correction of the eigenvalue ( $C^2\left(\left\vert\psi_0%
\right\rangle\right)$) can be given by $\frac{\left\langle
\psi_0^*\right\vert SM\left\vert\psi_0\right\rangle}{\left\langle
\psi_0^*\right\vert S\left\vert\psi_0\right\rangle}$. Note that $%
\left\vert\psi_k\right\rangle\sim[\varepsilon^{k},\varepsilon^{k+1},%
\varepsilon^{k+1},\varepsilon^{k+2}]^T$, $k=0,1$ and the matrix $S$ is
anti-diagonal. One can easily find that $\frac{\left\langle\psi_0^*\right%
\vert S M\left
\vert\psi_0\right\rangle}{\left\langle \psi_0^*\right\vert
S\left\vert\psi_0\right\rangle}\sim\varepsilon^6 $. Thus to a good
approximation, the eigenvalue of $\rho S\rho^*S$ is well determined by $%
\left\vert\psi_0\right\rangle\left\langle\psi_0\right\vert
S\left\vert\psi^*_0\right\rangle\left\langle\psi^*_0\right\vert S$ which
means the concurrence reads ( $\bar{A}_{0gg}\rightarrow 1$)%
\begin{equation}
C\left(\rho_{AB}\right)=C\left(\left\vert\psi_{0}\right\rangle\right)=2\left%
\vert \bar{A}_{0ee}-\bar{A}_{0ge}\bar{A}_{0eg}\right\vert.
\end{equation}

\section*{Acknowledgements (not compulsory)}

The idea is generated by YCS during his visiting Beijing Computational Science Research Center. YCS thanks C. P. Sun for his comments.
This work was supported by the National Natural Science Foundation of China,
under Grant No.11375036 and 11175033, the Xinghai Scholar Cultivation Plan
and the Fundamental Research Funds for the Central Universities under Grants
No. DUT15LK35 and No. DUT15TD47.

\section*{Author contributions statement}

YCS raised the question. ZY, ZJ and YCS contributed to the theoretical
procedure and wrote the paper.

\section*{Additional information}

Competing financial interests: The authors declare no competing financial
interests.

\end{document}